\documentclass[11pt,a4paper]{article}

\usepackage{amsmath}
\usepackage{amssymb}
\usepackage{graphicx}
\usepackage{xcolor}
\usepackage{hyperref}
\usepackage{bm}
\usepackage{a4wide}

\hypersetup{
    colorlinks,
    linkcolor={red!50!black},
    citecolor={blue!50!black},
    urlcolor={blue!80!black}
}

\definecolor{scipostdeepblue}{HTML}{002B49}

\DeclareSymbolFont{usualmathcal}{OMS}{cmsy}{m}{n}
\DeclareSymbolFontAlphabet{\mathcal}{usualmathcal}

\definecolor{scipostdeepblue}{HTML}{002B49}

\begin{document}

\begin{center}{\Large \textbf{\color{scipostdeepblue}{
Scar subspaces stabilized by algebraic closure:
Beyond equally-spaced spectra and exact solvability\\
}}}\end{center}

\begin{center}\textbf{
Chihiro Matsui\textsuperscript{1$\star$}
}\end{center}

\begin{center}
{\bf 1} Department of Mathematical Sciences, The University of Tokyo, 3-8-1 Komaba, Meguro-ku, Tokyo 153-8914, Japan
\\[\baselineskip]
$\star$ \href{mailto:matsui@ms.u-tokyo.ac.jp}{\small matsui@ms.u-tokyo.ac.jp}
\end{center}

\section*{\color{scipostdeepblue}{Abstract}}
\textbf{\boldmath{
We propose algebraic closure as a stabilizing mechanism for quantum many-body scar subspaces, showing that scar states need not rely on an embedded equally spaced spectrum or analytical solvability. We formulate Lie-algebraically closed invariant subspaces and construct local quantum many-body Hamiltonians that preserve an $\mathfrak{su}(N)$-invariant scar subspace but need not act diagonally within it. Such non-diagonal action mixes the conventional tower states while preserving the multidirectional lattice-like spectral structure, allowing nonthermal dynamics and persistent multifrequency oscillations to survive even when individual scar eigenstates become analytically intractable. 
}}

\vspace{\baselineskip}



\vspace{10pt}
\noindent\rule{\textwidth}{1pt}
\tableofcontents
\noindent\rule{\textwidth}{1pt}
\vspace{10pt}


\section{Introduction}
\label{sec:intro}

Quantum many-body scars (QMBS)~\cite{bib:BSKLOPCZEGVL17,bib:VRB17,bib:TMASP18,bib:TMASP18-2,bib:MRBR18,bib:MRB18,bib:MRB20,bib:MBJ20,bib:SYK20,bib:MBR22,bib:CIKM23,bib:B23,bib:M24} have attracted significant attention as a mechanism for weak ergodicity breaking in nonintegrable systems. In many known examples~\cite{bib:MRB18,bib:MRBR18,bib:SI19,bib:MRB20,bib:IS20,bib:SYK20,bib:CIKM23,bib:OM23}, QMBS are constructed on top of a solvable reference state and form an invariant subspace generated by a restricted spectrum-generating algebra (rSGA)~\cite{bib:MRB20}, leading to a tower of states with an equally spaced energy spectrum within the subspace. This structure is often associated with an emergent $\mathfrak{su}(2)$ algebra acting within the subspace~\cite{bib:CTPHMPSLA19,bib:IS20,bib:MM20,bib:OBCK20,bib:RLF21,bib:RLF22,bib:CIKM23,bib:WZGZ24} and underlies the characteristic periodic revivals observed in dynamics~\cite{bib:TMASP18,bib:TMASP18-2,bib:CIKM23}.  

Within the conventional tower-based paradigm, a remarkable feature is the apparent robustness of both partial solvability and equally spaced spectra within the invariant subspace. Indeed, it has been argued that locality enforces the persistence of equally spaced towers of solvable states, provided that the original rSGA-based tower states remain energy eigenstates~\cite{bib:OGMM26}. On the other hand, Lie-algebraic scar constructions are not restricted to a single $\mathfrak{su}(2)$ tower~\cite{bib:OBCK20,bib:O26}. Higher-rank Lie algebras provide multiple independent ladder directions, generating families of exact scar states with different energy increments and leading naturally to non-equally-spaced lattice spectra and multifrequency dynamics~\cite{bib:OBCK20}. Thus, equally spaced spectra are not an essential feature of Lie-algebraic scar constructions, despite their robustness within the conventional single-tower setting. These observations motivate a more fundamental question: what is the essential algebraic structure that stabilizes a scar subspace?

In this work, we address this question by shifting the focus from individual scar towers to the algebraic closure of the entire invariant subspace. We regard the invariant subspace $W$, rather than a particular tower basis within it, as the fundamental object. If the Hamiltonian acts within $W$ through a closed Lie-algebraic structure, the invariance of $W$ can be maintained even when the conventional tower states do not diagonalize the Hamiltonian. From this viewpoint, the familiar exactly solvable tower construction corresponds to a special case in which the restricted Hamiltonian is diagonal in a weight basis. More generally, the Hamiltonian may act non-diagonally within $W$, mixing the conventional tower states while preserving the algebraically closed subspace as a whole. Algebraic closure therefore provides a framework for stabilizing scar subspaces without requiring either a single equally spaced spectrum or an analytically solvable eigenstate basis. 

To formulate this viewpoint systematically, we first construct Lie-algebraically closed subspaces for general finite-dimensional highest-weight representations and then discuss embedded $\mathfrak{su}(2)$ and fundamental $\mathfrak{su}(N)$ realizations. 
We introduce local tower-mixing perturbations that act nontrivially within the resulting invariant subspace. Tower mixing changes the original tower states in both the $\mathfrak{su}(2)$ and higher-rank cases while preserving the algebraically closed invariant subspace. 

Within such an algebraically stabilized subspace, the Hamiltonian acts as a global representation of the underlying Lie algebra. Consequently, its spectral structure can be determined at the level of the Lie-algebraic representation without explicitly solving for the individual eigenstates or eigenenergies. 
In the $\mathfrak{su}(2)$ case, the mixed Hamiltonian can be reduced by an $\mathrm{SU}(2)$ rotation to a new equally spaced tower with a modified energy spacing. In the higher-rank case, the corresponding rotation still yields a multidirectional lattice-like spectrum, whose energy spacings and rotated HWS are generally not analytically solvable for $N \geq 5$. 
As a concrete demonstration, we construct a local $\mathfrak{su}(3)$ model and investigate observable signatures of the algebraically stabilized scar subspace, including anomalously low entanglement that persists under tower mixing and persistent single- and multifrequency revival dynamics. 

The rest of this paper is organized as follows. 
In Sec.~\ref{sec:suN}, we formulate Lie-algebraically closed subspaces for a general highest-weight representation and discuss the embedded $\mathfrak{su}(2)$ and fundamental $\mathfrak{su}(N)$ realizations. 
In Sec.~\ref{sec:tower_mixing}, we introduce tower-mixing perturbations and discuss their consequences for the invariant subspace and its spectral structure. 
In Sec.~\ref{sec:numerics}, we discuss the entanglement structure and revival dynamics. 
Finally, Sec.~\ref{sec:conclusion} summarizes our results and discusses future directions.

\section{Lie-algebraically closed subspace}
\label{sec:suN}

A scar subspace often appears as an invariant subspace $W$ that is not mixed with its complement under time evolution~\cite{bib:SAP21}. In many scar models, this closure is guaranteed by an additional mathematical structure emerging inside the scar subspace. The best-known example is a restricted spectrum-generating algebra (rSGA)~\cite{bib:MRB20}, which relates the Hamiltonian to a ladder operator within $W$ as
\begin{align}
    \big( [H,\,Q^{\dag}] - \mathcal{E}Q^{\dag} \big) \big|_W = 0.    
\end{align}
In many examples~\cite{bib:SI19,bib:MRB20,bib:OBCK20}, such an emergent extra algebraic relation reflects an underlying Lie algebraic structure within $W$. 

In this section, we provide a basic construction of a Lie-algebraically stabilized subspace, which serves as the platform for the scar-subspace construction discussed below. Importantly, algebraic closure itself does not require the reference state, or HWS, to admit an explicit analytical expression.

\subsection{General construction}

Throughout this paper, we consider models defined on a homogeneous tensor product $\mathcal{H} = V^{\otimes L}$ of a local Hilbert space $V$, where $V$ is a finite-dimensional local Hilbert space. 
Let $\mathfrak{g}$ be a Lie algebra and $\rho$ be its representation $\rho:\,\mathfrak{g} \to {\rm End}(V)$. The Lie algebra $\mathfrak{g}$ admits a natural extension to the tensor product space $V \otimes V$ through the coproduct 
\begin{align}
    \Delta(X) = X \otimes \bm{1} + \bm{1} \otimes X,
    \qquad X\in\mathfrak{g},
\end{align}
whose representation on $V\otimes V$ is 
\begin{align}
    (\rho\otimes\rho)\Delta(X)
    =\rho(X)\otimes\bm{1}+\bm{1}\otimes\rho(X). 
\end{align}
Since the coproduct satisfies coassociativity $(\Delta \otimes \bm{1}) \circ \Delta = (\bm{1} \otimes \Delta) \circ \Delta$, the local representation $\rho(\mathfrak{g})$ is consistently extended to the $L$-fold tensor product space $V^{\otimes L}$ as
\begin{align} \label{eq:coprod}
    \rho^{(L)}(X)
    := \rho^{\otimes L}(\Delta^{(L)}(X))
    =\sum_{j=1}^L \rho_j(X),
    \qquad X\in\mathfrak{g}. 
\end{align}

In this paper, we focus on a finite-dimensional highest-weight representation $\rho$. 
Let $|\lambda\rangle \in V$ be its highest-weight state (HWS) with highest weight $\lambda$. The direct-product state 
\begin{align}
    |\Psi_0\rangle = |\lambda\rangle^{\otimes L} \in V^{\otimes L}
\end{align}
is then a highest-weight state of the $L$-fold tensor-product representation with highest weight $L\lambda$. 

As a subspace stabilized by Lie-algebraic closure, we are particularly interested in the irreducible subspace $W_{\mathfrak{g}}$ with highest weight $L\lambda$, generated from $|\Psi_0 \rangle$ under the coproduct action of $\mathfrak{g}$, 
\begin{align}
    W_{\mathfrak{g}} := {\rm span}\{ \rho^{(L)}(X_1)\cdots \rho^{(L)}(X_m) |\Psi_0\rangle \,|\,X_1,\ldots,X_m\in\mathfrak{g} \}. 
\end{align}
Such a subspace is contained in the totally symmetric subspace of $V^{\otimes L}$. Indeed, since the reference state $|\Psi_0 \rangle = |\lambda \rangle^{\otimes L}$ is invariant under any permutation of sites and the coproduct action $\rho^{(L)}(X) = \sum_{j=1}^L \rho_j(X)$ commutes with all site permutations, every state generated from $|\Psi_0 \rangle$ remains totally symmetric, leading to $W_{\mathfrak{g}} \subseteq {\rm Sym}^L(V)$.

\subsection{Embedded $\mathfrak{su}(2)$ realization}
\label{sec:XY}

For simplicity, we hereafter denote the globally represented generator $\rho^{(L)}(X)$ simply by $X$, and its local action $\rho_j(X)$ by $X_j$. 

We first illustrate the general construction above for $\mathfrak{g}=\mathfrak{su}(2)$. In the construction presented here, $\mathfrak{su}(2)$ is embedded in the three-dimensional local Hilbert space $V \simeq \mathbb{C}^3 = {\rm span}\{|+ \rangle, |0 \rangle, |- \rangle \}$ of a physical spin-$1$ degree of freedom. 
We denote its Cartan generator by $h$ and the raising and lowering generators by $e$ and $f$, respectively, which satisfy
\begin{align}
    [h,e]=2e, \qquad
    [h,f]=-2f, \qquad
    [e,f]=h,
\end{align}
with $h^\dagger=h$ and $e^\dagger=f$.

For the embedded representation considered here, the local $\mathfrak{su}(2)$ action is closed within ${\rm span}\{|+\rangle,|-\rangle\}\subsetneq V$, while $|0\rangle$ lies outside this irreducible component. Starting from the highest-weight product state $|\Psi_0\rangle=|+\rangle^{\otimes L}$, repeated action of the global lowering operator generates
\begin{align}
    W_{\mathfrak{su}(2)}
    &= {\rm span}\{ f^n|\Psi_0\rangle \}_{n=0}^{L} \nonumber\\
    &= {\rm Sym}^L\left( {\rm span}\{|+\rangle,|-\rangle\} \right) \simeq {\rm Sym}^L(\mathbb{C}^2).
\end{align}
The second equation can be proved as follows. Since the global lowering operator is given by $f = \sum_{j=1}^L f_j$, the state $f^n|\Psi_0\rangle$ is, up to normalization, the totally symmetric state containing $n$ local states $|- \rangle$ and $L - n$ local states $|+ \rangle$. As $n$ runs from $0$ to $L$, these orthogonal $L + 1$ states form a basis of the totally symmetric subspace of ${\rm span}\{|+\rangle,|-\rangle\}^{\otimes L}$. Hence, we have ${\rm span}\{f^n|\Psi_0\rangle\}_{n=0}^L = {\rm Sym}^L\!\left({\rm span}\{|+\rangle,|-\rangle\}\right)$.

Note that, although $W_{\mathfrak{su}(2)}$ is contained in the totally symmetric subspace ${\rm Sym}^L(V)$ as discussed in the previous subsection, the inclusion is strict for the present embedding, $W_{\mathfrak{su}(2)} \subsetneq {\rm Sym}^L(\mathbb{C}^3)$. 
States in ${\rm Sym}^L(V)$ outside $W_{\mathfrak{su}(2)}$ are not protected by this construction and generically undergo thermalizing dynamics. 

The spin-$1$ $XY$ model in a uniform magnetization~\cite{bib:SI19} provides one of the simplest examples fitting this framework. The Hamiltonian  
\begin{align} \label{eq:H_XY}
    &H_{XY} = H_{\rm ann} + H_{\rm tower}, \\
    &H_{\rm ann} = \sum_{j=1}^L \left( S_j^x S_{j+1}^x + S_j^y S_{j+1}^y \right), \quad
    H_{\rm tower} = \frac{\mathcal{E}}{2} \sum_{j=1}^L S^z_j.  \nonumber
\end{align}
has an invariant scar subspace spanned by exactly solvable eigenstates $|\Psi^{\pm}_n\rangle$~\cite{bib:SI19}: 
\begin{align}
    |\Psi^{\pm}_n\rangle = (Q^{\pm})^n |\Psi^\pm_0\rangle,
    \qquad
    Q^{\pm} = \sum_{j=1}^L (-1)^j (S_j^{\pm})^2.   
\end{align} 
where $|\Psi^\pm_0\rangle$ are the fully polarized states, $S^\pm=S^x\pm iS^y$, and we take $L$ to be even. Throughout this paper, we impose periodic boundary conditions, with site indices understood modulo $L$.  

The annihilation part of the Hamiltonian vanishes upon restriction to the subspace $W_{XY}$ as $H_{\rm ann}W_{XY}=0$, while $H_{\rm tower}$ acts as the Cartan generator within the same subspace. Accordingly, the restricted operators satisfy the closed algebraic relations within $W_{XY}$~\cite{bib:SI19} 
\begin{align}
    &\Big( [H_{XY},\,Q^{\pm}] \mp \mathcal{E} Q^{\pm} \Big)\Big|_{W_{XY}} = 0, \\
    &\Big( [Q^+,\, Q^-] - \frac{\mathcal{E}}{2} H_{XY} \Big)\Big|_{W_{XY}} = 0. \nonumber
\end{align}
Thus, $W_{XY}$ realizes a twisted $\mathfrak{su}(2)$ structure (see Appendix~\ref{app:sym}). 
The first relation is an rSGA~\cite{bib:MRB20}, leading to the equally spaced tower spectrum and the associated single-frequency revivals. Importantly, $H_{XY}$ itself is not $\mathfrak{su}(2)$ invariant; the algebraic structure emerges only within the invariant subspace.

\subsection{Fundamental $\mathfrak{su}(N)$ realization}

In contrast to the embedded $\mathfrak{su}(2)$ realization above, we now consider the fundamental representation of $\mathfrak{su}(N)$ on the full local Hilbert space $V \simeq \mathbb{C}^N$. 
The Lie algebra $\mathfrak{su}(N)$ is generated by the Chevalley generators $\{e_a,f_a,h_a\}_{a=1,2,\dots,N-1}$, together with the composite root generators, subject to the Hermiticity conditions $h_a^\dagger=h_a$ and
$e_a^\dagger=f_a$. The defining relations are
\begin{align} \label{eq:suN}
    &[h_a,\,h_b]=0,\\
    &[h_a,\,e_b]=A_{ab}e_b,\quad
    [h_a,\,f_b]=-A_{ab}f_b,\nonumber\\
    &[e_a,\,f_b]=\delta_{ab}h_b,\nonumber
\end{align}
for $a,b=1,2,\dots,N-1$, where $A_{ab}$ is the Cartan matrix~\cite{bib:Humphreys,bib:Hall}. 

Let $|\lambda\rangle\in V$ be the highest-weight state of the fundamental representation. Starting again from the product HWS
\begin{align} \label{eq:reference}
    |\Psi_{\bm 0}\rangle:=|\lambda\rangle^{\otimes L} \in V^{\otimes L},
\end{align}
the global $\mathfrak{su}(N)$ action generates the irreducible subspace $W_{\mathfrak{su}(N)}$ with highest weight $L\lambda$.
Since 
\begin{align}
    &e_a|\Psi_{\bm0}\rangle=0,
    \quad a=1,2,\dots,N-1,\nonumber\\
    &f_a|\Psi_{\bm0}\rangle=0,
    \quad a=2,3,\dots,N-1,
\end{align}
we choose the ordered representatives
\begin{align} \label{eq:PBW}
    |\Psi_{\bm n}\rangle
    &= f_{N-1}^{n_{N-1}}\cdots f_2^{n_2}f_1^{n_1} |\Psi_{\bm0}\rangle
    =: \overleftarrow{\prod}_{a=1}^{N-1} f_a^{n_a}|\Psi_{\bm0}\rangle,
\end{align}
with $0\leq n_{N-1}\leq\cdots\leq n_2\leq n_1\leq L$. Here $\bm{n} =(n_1,n_2,\dots,n_{N-1})$ labels the numbers of simple lowering operations in the ordered representative. 

These ordered states $|\Psi_{\bm n}\rangle$ span the totally symmetric subspace. By introducing the occupation numbers as 
\begin{align}
    N_1 &= L-n_1, \\
    N_a &= n_{a-1}-n_a,
    \qquad a=2,\dots,N-1,\nonumber \\
    N_N &= n_{N-1}, \nonumber
\end{align}
the conditions $0\leq n_{N-1} \leq \cdots \leq n_1 \leq L$ are equivalent to $N_a \geq 0$ and $\sum_{a=1}^N N_a = L$. 
Thus, $\bm{n}$ is in one-to-one correspondence with all occupation configurations of $L$ local states in the fundamental representation. 
For each occupation configuration, the state $|\Psi_{\bm n}\rangle$ is totally symmetric. Moreover, the states with different occupation configurations are mutually orthogonal. Therefore, these ordered states span totally symmetric subspace $W_{\mathfrak{su}(N)} = {\rm Sym}^L(V) \simeq {\rm Sym}^L(\mathbb{C}^N)$.

The totally symmetric structure of $W_{\mathfrak{su}(N)}$ also provides a simple local construction of a Hamiltonian that annihilates the entire subspace~\cite{bib:OBCK20}. 
Since $W_{\mathfrak{su}(N)} = {\rm Sym}^L(V)$, every nearest-neighbor pair is supported on the locally symmetric subspace ${\rm Sym}^2(V)$. Therefore, any local operator satisfying
\begin{align}
    h_{\rm ann}^{\mathfrak{su}(N)}\,{\rm Sym}^2(V)=0
\end{align}
gives a frustration-free annihilator 
\begin{align}\label{eq:ann_suN}
    H_{\rm ann}^{\mathfrak{su}(N)}
    =\sum_{j=1}^L (h_{\rm ann}^{\mathfrak{su}(N)})_{j,j+1},
    \qquad
    H_{\rm ann}^{\mathfrak{su}(N)}W_{\mathfrak{su}(N)} = 0.
\end{align}
Together with a Cartan term
\begin{align}\label{eq:suN_tower}
    H_{\rm tower}^{\mathfrak{su}(N)}
    = \sum_{j=1}^L \sum_{a=1}^{N-1} \lambda_a(h_a)_j,
\end{align}
this yields a local Hamiltonian $H^{\mathfrak{su}(N)}$ that preserves $W_{\mathfrak{su}(N)}$ while assigning different energy increments to different lowering directions within $W_{\mathfrak{su}(N)}$: 
\begin{align}
    &H^{\mathfrak{su}(N)} |\Psi_{\bm{n}} \rangle = H_{\rm tower}^{\mathfrak{su}(N)} |\Psi_{\bm{n}} \rangle = E_{\bm{n}} |\Psi_{\bm{n}} \rangle, \nonumber \\
    &E_{\bm{n}} = \sum_{a=1}^{N-1} \lambda_a \Big( L \delta_{a,1} - \sum_{b=1}^{N-1} A_{ab} n_b \Big). \label{eq:suN_energy}
\end{align}

\section{Tower mixing in scar subspaces stabilized by algebraic closure}
\label{sec:tower_mixing}

Since there is no universally accepted definition of QMBSs, we clarify
the terminology used in this work. We refer to eigenstates within an
invariant subspace whose dimension grows polynomially with system size
as QMBSs when their entanglement entropy obeys a sub-volume law and
they exhibit nonthermal behavior distinct from that expected for
generic many-body eigenstates. 
We will show below that the algebraically stabilized subspaces considered here satisfy these criteria.

The viewpoint of algebraic closure as a stabilizing mechanism of an invariant subspace provides a new perspective on perturbations that avoid complete thermalization. In particular, we consider perturbations that preserve the algebraically stabilized subspace introduced in the previous sections.

We focus on perturbations that preserve the scar subspace $W$ itself while mixing the tower states within $W$. As a result, the individual tower states need not remain eigenstates of the perturbed Hamiltonian, whereas the algebraically closed scar subspace survives as an invariant subspace.

\subsection{Tower-mixing perturbation for the $\mathfrak{su}(2)$ case} \label{sec:rhombus}

In this subsection, we discuss the simplest example of the $\mathfrak{su}(2)$
case. We show that both the equally spaced scar tower and the analytic solvability of each tower state within the invariant subspace remain intact. 

The simplest tower-mixing perturbation of this type was originally considered in \cite{bib:CIKM23} in the form 
\begin{align}
    &H^{\mathfrak{su}(2)}_{\rm rhom}
    = m \sum_{j=1}^{L} (h^{\mathfrak{su}(2)}_{\rm rhom})_{j}, \qquad
    (h^{\mathfrak{su}(2)}_{\rm rhom})_{j}
    = \frac{(-1)^j}{2}\big((S_j^+)^2 + (S_j^-)^2\big). 
\end{align}
This perturbation mixes different states within the tower, and therefore the original tower states are no longer eigenstates of the full Hamiltonian $H_{XY} + H^{\mathfrak{su}(2)}_{\rm rhom}$. Nevertheless, the scar subspace $W_{\mathfrak{su}(2)} = {\rm span}\{|\Psi_n\rangle\}$ remains invariant by algebraic closure, since, when restricted to $W_{\mathfrak{su}(2)}$, both $H_{XY}$ and $H^{\mathfrak{su}(2)}_{\rm rhom}$ act as effective $\mathfrak{su}(2)$ generators. 

As a consequence, the effective Hamiltonian restricted to $W_{\mathfrak{su}(2)}$ is unitarily equivalent to a Cartan element of $\mathfrak{su}(2)$ by a uniform $\mathrm{SU}(2)$ rotation:
\begin{align}
    \left(H_{XY} + H^{\mathfrak{su}(2)}_{\rm rhom}\right)\Big|_{W_{\mathfrak{su}(2)}}
    = -\frac{1}{2}\sqrt{\mathcal{E}^2 + m^2} \sum_{j=1}^L \widetilde{S}^z_j,
\end{align}
where $\widetilde{S}_j^z = G S_j^z G^\dagger$ for all $j$ ($G \in \mathrm{SU}(2)$). 

The resulting eigenstates form a rotated tower
\begin{align} \label{eq:rotated_tower}
    |\Psi^G_n\rangle = G^{\otimes L} |\Psi_n\rangle = (\widetilde{S}^-_{\rm tot})^n |\Psi^G_0\rangle, 
\end{align}
which is produced by the rotated ladder operators $\widetilde{S}^\pm_{\rm tot} := G^{\otimes L} S^\pm_{\rm tot} (G^\dagger)^{\otimes L}$. Although the original tower is mixed by an $\mathrm{SU}(2)$ rotation $G$, the eigenstates within $W_{\mathfrak{su}(2)}$ remain analytically accessible. 
As a result, the rSGA structure within $W_{\mathfrak{su}(2)}$ is preserved with a modified energy spacing: 
\begin{align}
    \Big(
    [H_{XY} + H^{\mathfrak{su}(2)}_{\rm rhom},\, \widetilde{S}^{\pm}_{\rm tot}]
    \pm \sqrt{\mathcal{E}^2 + m^2}\,\widetilde{S}^{\pm}_{\rm tot}
    \Big)\Big|_{W_{\mathfrak{su}(2)}} = 0. 
\end{align}
The perturbed model therefore still has an equally spaced energy spectrum within $W_{\mathfrak{su}(2)}$. 

This simple persistence of an equally spaced scar tower is special to the $\mathfrak{su}(2)$ case. In the higher-rank generalization discussed below, algebraic closure still protects the invariant subspace and its lattice-like spectral structure, while the analytical solvability of individual eigenstates is generally lost.

\subsection{Tower-mixing perturbation for the higher-rank case} \label{app:rhombus_suN}

A simple generalization of a tower-mixing perturbation to the higher-rank case is realized by the Hermitian linear combination of the $\mathfrak{su}(N)$ ladder operators
\begin{align} \label{eq:rhom_suN}
    &H^{\mathfrak{su}(N)}_{\rm rhom} = \sum_{j=1}^L (h_{\rm rhom}^{\mathfrak{su}(N)})_j, \qquad
    h_{\rm rhom}^{\mathfrak{su}(N)}
     = \sum_{1 \leq a<b \leq N} \big(m_{ab}\, e_{a,b} + m^*_{ab}\, f_{b,a} \big), 
\end{align}
for $m_{ab} \in \mathbb{C}$. Here we use the notation  
\begin{align}
    e_{a,b} = |a \rangle\langle b|, \qquad
    f_{b,a} = |b \rangle\langle a| 
\end{align}
for $1 \leq a<b \leq N$. 
The Chevalley generators are identified as $e_a = e_{a,a+1}$ and $f_a = f_{a+1,a}$ in this notation, while the remaining operators correspond to composite-root generators. 
The composite roots $e_{1,a}$ and $f_{a,1}$ ($a = 2,\dots,N$) are regarded as annihilation and creation operators of a particle of species $a$, respectively. 
For the numerical demonstrations in Sec.~\ref{sec:numerics}, we explicitly provide the Hamiltonian of the $N=3$ case. In the basis $\{|+\rangle,|0\rangle,|-\rangle\}$, the annihilation term is given by 
\begin{align} \label{eq:local_h}
    h^{\mathfrak{su}(3)}_{\rm ann} 
    &= a_+ \big(|+0\rangle \langle+0| - |+0\rangle \langle0+|  - |0+\rangle \langle+0| + |0+\rangle \langle0+| \big)  \\
    &+ a_- \big(|-0\rangle \langle-0| - |-0\rangle \langle0-| - |0-\rangle \langle-0| + |0-\rangle \langle0-| \big) \nonumber \\
    &+ a_0 \big(|+-\rangle \langle+-| -  |+-\rangle \langle-+| -  |-+\rangle \langle+-| + |-+\rangle \langle-+| \big), \nonumber
\end{align}
where $a_+$, $a_-$, and $a_0$ are real coefficients, the energy splitting term is given by
\begin{align} \label{eq:Zeeman}
    h^{\mathfrak{su}(3)}_{\rm tower} = \lambda_1 h_1 + \lambda_2 h_2 
\end{align}
with real $\lambda_1$ and $\lambda_2$, and the tower mixing perturbation is given by 
\begin{align} \label{eq:rhom_su3}
    h_{\rm rhom}^{\mathfrak{su}(3)}
    &= m_1 |+\rangle\langle 0| + m_1^* |0 \rangle\langle +| \nonumber \\
    &+ m_2 |0\rangle\langle -| + m_2^* |- \rangle\langle 0| \nonumber \\ 
    &+ m_3 |+\rangle\langle -| + m_3^* |- \rangle\langle +|  
\end{align}
with $m_1,m_2,m_3 \in \mathbb{C}$. 
This perturbation mixes different states within $W_{\mathfrak{su}(N)}$, and therefore the original tower states are no longer eigenstates of the full Hamiltonian $H^{\mathfrak{su}(N)} + H^{\mathfrak{su}(N)}_{\rm rhom}$. 

On the other hand, the subspace $W_{\mathfrak{su}(N)}$ itself remains invariant. Indeed, since $H^{\mathfrak{su}(N)}_{\rm ann}$ annihilates $W_{\mathfrak{su}(N)}$, the restricted action of the perturbed Hamiltonian is
\begin{align}
    \Big(H^{\mathfrak{su}(N)}
    + H^{\mathfrak{su}(N)}_{\rm rhom} \Big)\Big|_{W_{\mathfrak{su}(N)}} 
    = H^{\mathfrak{su}(N)}_{\rm tower}
    + H^{\mathfrak{su}(N)}_{\rm rhom}
    = \sum_{j=1}^L \big( h^{\mathfrak{su}(N)}_{\rm tower} + h^{\mathfrak{su}(N)}_{\rm rhom} \big)_j. 
\end{align}
The local terms $h^{\mathfrak{su}(N)}_{\rm tower}$, given by \eqref{eq:suN_tower}, and $h^{\mathfrak{su}(N)}_{\rm rhom}$, given by \eqref{eq:rhom_suN} are linear combinations of $\mathfrak{su}(N)$ generators, and therefore map $W_{\mathfrak{su}(N)}$ into itself by algebraic closure.

Since the restricted Hamiltonian is a translationally invariant on-site $\mathfrak{su}(N)$ element, it can be diagonalized into Cartan elements by a uniform $\mathrm{SU}(N)$ transformation $G^{\otimes L}$ with $G\in \mathrm{SU}(N)$:  
\begin{align}
    &\Big(H^{\mathfrak{su}(N)}
    + H^{\mathfrak{su}(N)}_{\rm rhom} \Big)\Big|_{W_{\mathfrak{su}(N)}}
    = \sum_{j=1}^L \big( h^{\mathfrak{su}(N)}_{\rm tower} + h^{\mathfrak{su}(N)}_{\rm rhom} \big)_j
    = \sum_{j=1}^L \sum_{a=1}^{N-1} \mu_a \big( \widetilde{h}_a \big)_j, \label{eq:suN_hG}
\end{align}
where $\mu_a$ are real coefficients and $\widetilde{h}_a := G h_a G^{\dag}$ ($a=1,2,\dots,N-1$) are the rotated Cartan generators. 

Correspondingly, the eigenstates within $W_{\mathfrak{su}(N)}$ can be
written as the tower states generated from the rotated HWS
\begin{align}
    &|\Psi_0^G \rangle
    = G^{\otimes L}|\Psi_0\rangle
    = \big(G|\lambda\rangle\big)^{\otimes L}, \\
    &|\Psi_{\bm n}^G\rangle
    = (f^G_{N-1})^{n_{N-1}} \cdots (f^G_1)^{n_1} |\Psi_0^G \rangle, \qquad
    f^G_a := G^{\otimes L} f_a (G^\dagger)^{\otimes L},
\end{align}
where $f^G_a := \sum_{j=1}^L G (f_a)_j G^{\dag}$ for $0 \leq n_{N-1} \leq \cdots \leq n_1 \leq L$. These states diagonalize the restricted Hamiltonian \eqref{eq:suN_hG}. 
Thus, the spectrum within the subspace $W_{\mathfrak{su}(N)}$ is deformed into another lattice with modified spacings,
\begin{align} \label{eq:energy_def}
    E^G_{\bm{n}} = \sum_{a=1}^{N-1} \mu_a \Big( L\delta_{a,1} - \sum_{b=1}^{N-1} A_{ab} n_b \Big).    
\end{align}
This shows that the lattice-like energy spectrum survives even after the tower-mixing perturbation. We emphasize that, while the rotated HWS $|\Psi_0^G \rangle=(G|\lambda \rangle)^{\otimes L}$ remains a direct product state, neither the coefficients $\mu_a$ nor the local state $G|\lambda \rangle$ can in general be expressed in closed form for generic $N \geq 5$.

In this way, under a tower-mixing perturbation in the $\mathfrak{su}(N)$ case, the original tower states are generally mixed and cease to be eigenstates, while the analytical solvability of the rotated HWS and the corresponding energy spacings is not guaranteed. Nevertheless, such perturbations preserve not only the invariant subspace $W_{\mathfrak{su}(N)}$ of the unperturbed Hamiltonian, but also its lattice-like spectral structure. Thus, neither closed-form energy spacings nor an analytically solvable HWS is required for the algebraic stabilization of the invariant subspace.

\section{Observable consequences}
\label{sec:numerics}

So far, we have discussed a class of Hamiltonians that admit invariant subspaces protected by algebraic closure. Such invariant subspaces consist of multidirectional towers of states associated with a Lie algebra, and exhibit a lattice-like energy spectral structure within the subspace. We have also shown that these invariant subspaces can survive under tower-mixing perturbations: the subspace remains protected by algebraic closure, although the individual tower states generally become analytically intractable. 

In this section, we demonstrate that such an invariant subspace can appear as a scar subspace by showing that the tower states have anomalously low entanglement entropy. We also confirm the observable consequence of the lattice-like energy spectrum through persistent oscillations. Unlike the $\mathfrak{su}(2)$ case, where single-frequency oscillations are observed, the $\mathfrak{su}(N)$ case exhibits multifrequency oscillations reflecting the multidirectional spectral structure. Importantly, these characteristic features persist even for analytically intractable tower states. 
We demonstrate these characteristic features in the $\mathfrak{su}(3)$ case as a representative example.

\subsection{Energy vs entanglement entropy}

In this subsection, we demonstrate that, for the local Hamiltonian defined by \eqref{eq:local_h}, \eqref{eq:Zeeman}, and \eqref{eq:rhom_su3}, the eigenstates in $W_{\mathfrak{su}(3)}$ exhibit anomalously low entanglement entropy, which is a characteristic signature of quantum many-body scars~\cite{bib:ZSMK21,bib:MBR22,bib:CIKM23}. These results demonstrate that, in the present local Hamiltonian, the algebraically closed invariant subspace is indeed realized as a nonthermal scar subspace.

We first consider the original states $|\Psi_{n_1,n_2} \rangle$ in $W_{\mathfrak{su}(3)}$, constructed on top of the fully-polarized reference state. Such states are uniquely labeled by a set of integers $(n_1,n_2)$, which are in one-to-one correspondence with the eigenvalues of the conserved quantities, the total magnetization $M:=\sum_j S^z_j$ and the total squared magnetization $Q:=\sum_j (S^z_j)^2$, of the Hamiltonian $H^{\mathfrak{su}(3)}$, via the relations
\begin{align} \label{eq:MQ_to_n}
    M_{n_1,n_2} = L - n_1 - n_2, \quad
    Q_{n_1,n_2} = L - n_1 + n_2. 
\end{align}

Since these tower states are generated by applying translationally invariant local operators on the direct product state \eqref{eq:PBW}, they admit matrix-product state (MPS) representations. In particular, for a tower state labeled by $\bm{n} = (n_1,\ldots,n_{N-1})$, the corresponding tower state has an MPS expression with a bond dimension given by $\prod_{a=1}^{N-1} (n_a+1)$ (see also Appendix~\ref{app:MPS}). 
Therefore, its entanglement entropy is bounded from above by the logarithm of its MPS bond dimension, $\log \prod_{a=1}^{N-1} (n_a+1) = \sum_{a=1}^{N-1} \log(n_a+1)$~\cite{bib:OBCK20}, which scales as $O(\log L)$ even in the finite-density regime and thus obeys a sub-volume law.

\begin{figure*}[t]
    \centering
    \includegraphics[width=1.0\linewidth]{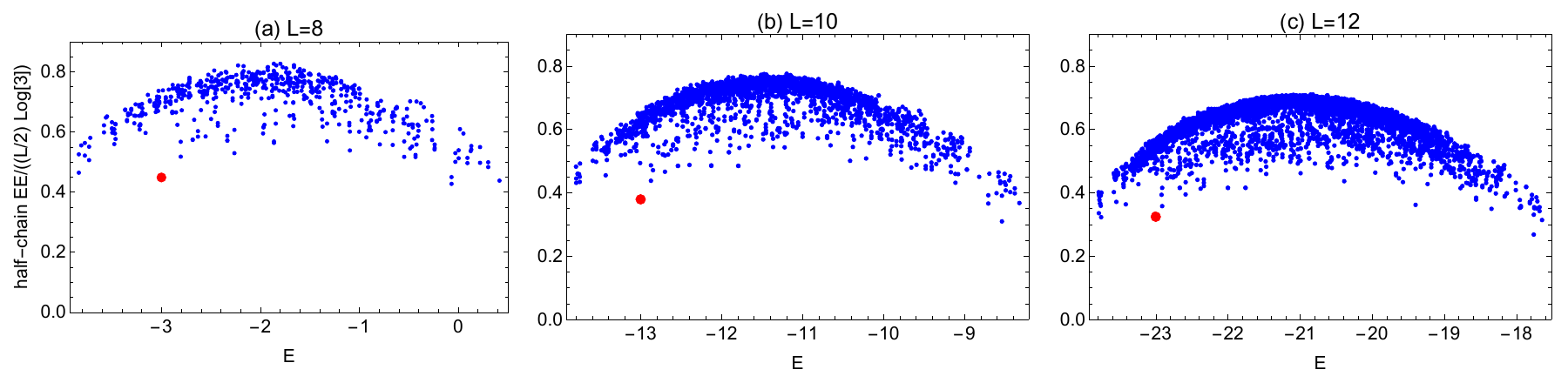}
    \caption{
    Normalized half-chain von Neumann entanglement entropy versus energy in the sector labeled by $(M_{n_1,n_2}^G,Q_{n_1,n_2}^G) = (1,5)$ for system sizes (a) $L=8$, (b) $L=10$, and (c) $L=12$. The red dots indicate eigenstates within $W_{\mathfrak{su}(3)}$. 
    Parameters: $(\lambda_1,\lambda_2) = (5/2,-5/2)$; $(a_+,a_0,a_-) = (1/3,-1/5,1/3)$; $(m_1,m_2,m_3) = (0,0,1/2)$. }
     \label{fig:EE}
\end{figure*}

The low-entanglement property is preserved even after a translationally invariant tower-mixing perturbation. As shown in the previous section, the perturbed Hamiltonian $H^{\mathfrak{su}(3)} + H^{\mathfrak{su}(3)}_{\rm rhom}$ reduces to a sum of Cartan generators by a uniform tensor product of on-site $\mathrm{SU}(3)$. 
Since an on-site unitary transformation does not change the bipartite entanglement entropy, the entanglement structure within $W_{\mathfrak{su}(3)}$ remains unchanged from that of the unperturbed case. 

The same argument applies to the higher-rank case: any translationally invariant tower-mixing perturbation given by a local $\mathfrak{su}(N)$ element can be diagonalized into Cartan generators by a uniform tensor product of on-site $\mathrm{SU}(N)$ rotations. This confirms that low entanglement entropy is a consequence of algebraic closure of the subspace $W_{\mathfrak{su}(N)}$ realized by a local Hamiltonian, and therefore is independent of solvability of individual states (see Appendix~\ref{app:MPS}). 

Figure~\ref{fig:EE} shows the half-chain von Neumann entanglement entropy in the symmetry sector  $(M_{n_1,n_2}^G,Q_{n_1,n_2}^G) = (1,5)$. For the parameters used in Fig.~\ref{fig:EE}, the Hamiltonian is related to the unperturbed case by the uniform rotation
\begin{align}
    G=
    \begin{pmatrix}
     1/\sqrt{2} & 0 & 1/\sqrt{2}\\
      0 & 1 & 0\\
     -1/\sqrt{2} & 0 & 1/\sqrt{2}
    \end{pmatrix}
    \in \mathrm{SU}(3),
\end{align}
under which the conserved quantities are transformed as $M^G = G^{\otimes L} M (G^\dagger)^{\otimes L}$ and $Q^G = G^{\otimes L} Q (G^\dagger)^{\otimes L}$. 
The eigenstates in $W_{\mathfrak{su}(3)}$ (red points) lie in the middle of the spectrum yet exhibit anomalously low entanglement entropy, clearly separated from surrounding eigenstates not belonging to $W_{\mathfrak{su}(3)}$.

\subsection{Multifrequency revival dynamics}

A model with a small invariant subspace protected by a higher-rank Lie algebra exhibits revival behavior that is qualitatively distinct from that of conventional rSGA-induced scar models associated with $\mathfrak{su}(2)$. The time evolution of a physical quantity depends on how the initial state overlaps with the algebraically protected invariant subspace, or equivalently, on how many independent ladder directions are involved in the initial state. 

\begin{figure*}[t]
    \centering
    \includegraphics[width=0.25\textwidth]{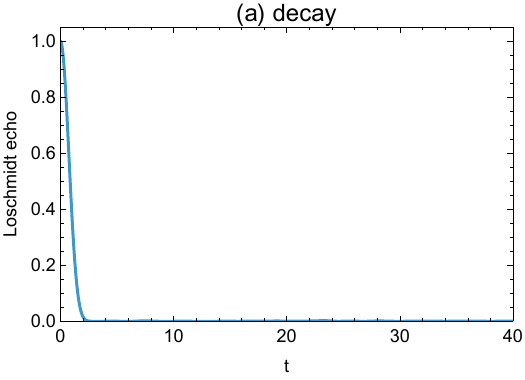}
    \hfill
    \includegraphics[width=0.24\textwidth]{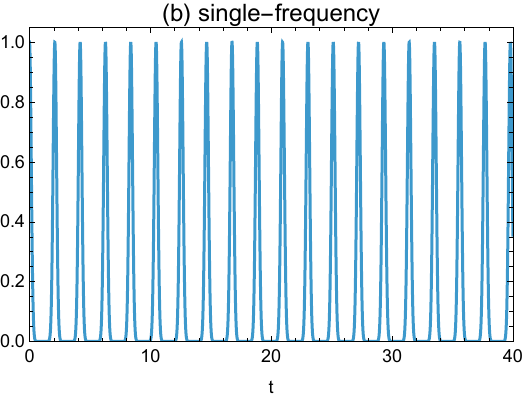}
    \hfill
    \includegraphics[width=0.24\textwidth]{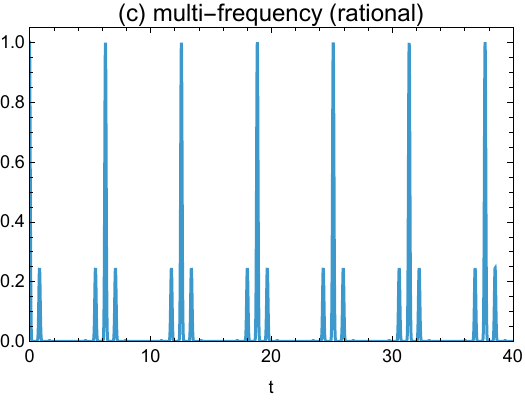}
    \hfill
    \includegraphics[width=0.24\textwidth]{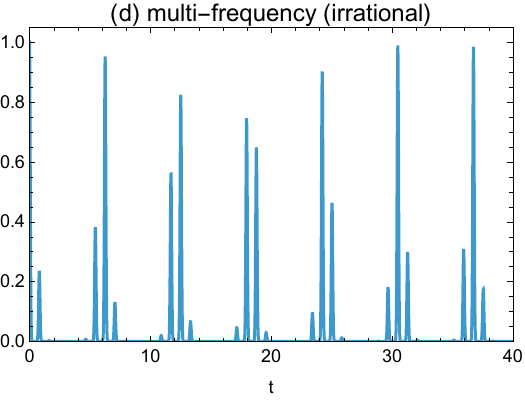}
    \caption{
    Loschmidt echo ($L=12$). 
    (a) rapid decay, (b) single-frequency revival, (c) multifrequency revival with rational $\omega_1/\omega_2$, and (d) multifrequency revival with irrational $\widetilde{\omega}_1/\widetilde{\omega}_2$. 
    Parameters: $(\lambda_1,\lambda_2)=(2,-3)$ for (a),(c),(d) and $(1,2)$ for (b); $(a_+,a_0,a_-)=(1/3,-1/5,1/3)$; $(m_1,m_2,m_3)=(0.2,0.3,0.1)$ for (d) and $(0,0,0)$ otherwise; 
    Initial states: $|+0+-\rangle^{\otimes L/4}$ for (a) and $\otimes_j (|+ \rangle + |0\rangle + |-\rangle)/\sqrt{3}$ otherwise. }
     \label{fig:echo}
\end{figure*}

The revival dynamics is governed by the lattice structure of the energy spectrum \eqref{eq:suN_energy}. For $N=3$, the Loschmidt echo exhibits rapid decay when an initial state only has a small overlap with $W_{\mathfrak{su}(3)}$ (Fig.~\ref{fig:echo} (a)). On the other hand, for a state with large overlap with $W_{\mathfrak{su}(3)}$, it displays oscillations with frequencies given by energy differences $E_{n_1,n_2} - E_{n'_1,n'_2}$ (Fig.~\ref{fig:echo} (b)-(d)).

When the scar subspace consists of the eigenstates on top of the fully-polarized reference state \eqref{eq:PBW}, the frequencies are given in terms of two fundamental frequencies $\omega_1 = -2\lambda_1 + \lambda_2$ and $\omega_2 = \lambda_1 - 2\lambda_2$, where $\lambda_1$ and $\lambda_2$ are the Cartan coefficients in \eqref{eq:Zeeman}. 

When the ratio $\omega_1 / \omega_2$ is rational, the dynamics exhibits periodic revivals, as shown in Fig.~\ref{fig:echo} (b),(c). 
A single-frequency oscillation, shown in Fig.~\ref{fig:echo} (b), is recovered when the $\mathfrak{su}(3)$ structure effectively reduces to an $\mathfrak{su}(2)$, by activating only a single-directional tower. Such single-directional tower states span an effectively $\mathfrak{su}(2)$-invariant subspace, and the rSGA emerges within this subspace, explaining the single-frequency revivals.

As shown in the previous section, even when the tower states are mixed by a perturbation, algebraic closure ensures that the energy lattice structure within the invariant subspace remains intact up to local unitary group rotations \eqref{eq:energy_def}. 

Figure~\ref{fig:echo}(d) shows that multifrequency revivals persist even in the presence of the tower-mixing perturbation \eqref{eq:rhom_su3}. 
In this case, the original tower states are no longer eigenstates of the perturbed Hamiltonian. Nevertheless, the oscillation frequencies are still determined by energy differences on the rotated lattice, $\widetilde{\omega}_1 = -2\mu_1 + \mu_2$, $\widetilde{\omega}_2 = \mu_1 - 2\mu_2$, where $\mu_1$ and $\mu_2$ are the rotated Cartan coefficients defined in \eqref{eq:suN_hG}. 
When the ratio $\widetilde{\omega}_1/\widetilde{\omega}_2$ is irrational, the revival dynamics becomes quasiperiodic, as seen in Fig.~\ref{fig:echo}(d). 

In this way, multifrequency revivals in physical observables are a consequence of the multidirectional lattice-like energy spectrum, which survives even after tower-mixing perturbations. In the higher-rank case, a tower-mixing perturbation generally makes the individual tower states analytically intractable, while preserving the lattice-like energy spectral structure. This confirms that persistent multifrequency revivals are controlled by the algebraic closure that stabilizes the scar subspace, rather than by exact solvability of individual scar states.

\section{Conclusion}
\label{sec:conclusion}

In this work, we have proposed that neither equally spaced spectra nor exact solvability of individual eigenstates are fundamental requirements for the emergence and stability of scar subspaces. 

We first formulated Lie-algebraically closed subspaces for general
finite-dimensional highest-weight representations and constructed a
local Hamiltonian with an $\mathfrak{su}(N)$-invariant subspace using
local constraints on the symmetric subspace. In contrast to the familiar $\mathfrak{su}(2)$ single-tower
case, the invariant subspace of the Hamiltonian is organized into
multidirectional towers parametrized by multiple quantum numbers,
leading to a lattice-like energy structure rather than a single
equally spaced ladder.

We then considered local tower-mixing perturbations that act
nontrivially within the algebraically closed subspace. Although such
perturbations mix the original tower states and cause them to cease to
be eigenstates, the invariant subspace itself remains exactly
preserved. In the $\mathfrak{su}(2)$ case, the restricted Hamiltonian
can be reduced by an $\mathrm{SU}(2)$ rotation to a new equally spaced
tower with a modified energy spacing. In the higher-rank case, the
corresponding rotation preserves a multidirectional lattice-like
spectral structure, while neither the resulting energy spacings nor
the rotated HWS need to be analytically solvable. This illustrates that
analytical solvability of the reference state or individual scar
eigenstates is not essential to the algebraic stabilization mechanism.

We further showed analytically that the low-entanglement structure is
preserved under tower mixing: all eigenstates within the
$\mathfrak{su}(N)$-invariant subspace are related by a uniform on-site
$\mathrm{SU}(N)$ rotation to the original tower states and therefore
retain the same sub-volume-law entanglement bound. As a concrete
realization, we studied a local $\mathfrak{su}(3)$ model, for which
numerical results up to $L=12$ illustrate the anomalously low
entanglement of these scar eigenstates relative to surrounding
eigenstates. The corresponding Loschmidt-echo dynamics further
illustrates persistent single- and multifrequency oscillations
associated with different directions of the lattice-like spectrum,
including in the presence of tower mixing.

Our results therefore suggest that the essential role of the
Lie-algebraic structure is more naturally understood through closure
of the invariant subspace than through a particular solvable tower
basis. This perspective separates the exact preservation of the scar
subspace from the stronger requirement of analytical solvability of
its individual eigenstates, and provides a systematic setting for
constructing scar subspaces beyond conventional equally spaced,
exactly solvable towers.

The present framework opens several directions for future work,
including extensions to $q$-deformed algebras, the incorporation of
site-dependent perturbations, and the design of tunable dynamical
waveforms based on multidirectional spectra.
Another important direction is to extend the construction to scar
subspaces built on correlated reference states, such as matrix-product
states, which may provide a bridge to symmetry-protected topological
structures.

\section*{Acknowledgements}
C. M. thanks N. O'Dea and H. Katsura for helpful comments. This work was supported by JSPS KAKENHI Grant Nos. JP23K03244 and JP26K06937.

\begin{appendix}
\numberwithin{equation}{section}

\section{Algebraic closure under momentum-like phase twists}
\label{app:sym}

In this appendix, we consider a class of phase-twisted realizations of the algebraically closed subspace $W_{\mathfrak{su}(N)}$. We focus on momentum-like twists, 
\begin{align} \label{eq:twist}
    &\Delta_U^{(L)}(f_a)
    := U f_a U^\dagger
    = \sum_{j=1}^L e^{i(k_a-k_{a-1})j}(f_a)_j, \nonumber\\
    &\Delta_U^{(L)}(e_a)
    := U e_a U^\dagger
    = \sum_{j=1}^L e^{-i(k_a-k_{a-1})j}(e_a)_j,
\end{align}
for $a = 1, \dots, N-1$, in which the phase of each local weight state varies linearly with the site index. This class is particularly natural for translationally invariant systems, since it maps the uniform global root generators to momentum-carrying ones while preserving their $\mathfrak{su}(N)$ algebraic closure. 

Such a transformation is realized by the site-dependent on-site unitary
transformation
\begin{align}
    U=\bigotimes_{j=1}^L U_j,\qquad
    U_j={\rm diag} \left( 1,e^{ik_1j},e^{ik_2j},\dots,e^{ik_{N-1}j} \right),
\end{align}
where we set $k_0:=0$. 
Since the transformed generators \eqref{eq:twist} are related to the original ones by a unitary transformation, they obey the same $\mathfrak{su}(N)$ commutation relations. 

The corresponding invariant subspace is defined by 
\begin{align}
    W_{\mathfrak{su}(N)}^U := U W_{\mathfrak{su}(N)} = U\,{\rm Sym}^L(V), 
\end{align}
and can be regarded as a twisted symmetric subspace. Thus, the phase twist preserves the algebraic closure, while mapping the original invariant subspace to its twisted counterpart. 

The local annihilation constraints are transformed accordingly. Since every two-site component of $W_{\mathfrak{su}(N)}$ is supported on ${\rm Sym}^2(V)$, the corresponding local subspace after the twist is ${\rm Sym}^2_U(V)_{j,j+1} := (U_j\otimes U_{j+1}){\rm Sym}^2(V)$. 
The twisted annihilation term 
\begin{align}
    (h_{\rm ann}^{\mathfrak{su}(N),U})_{j,j+1}
    &:=
    (U_j\otimes U_{j+1})
    (h^{\mathfrak{su}(N)}_{\rm ann})_{j,j+1}
    (U_j^\dagger\otimes U_{j+1}^\dagger),
\end{align}
then locally annihilates the twisted symmetric space 
\begin{align}
    (h_{\rm ann}^{\mathfrak{su}(N),U})_{j,j+1} {\rm Sym}^2_U(V)_{j,j+1} = 0
\end{align}
for every $j$. Thus, we obtain 
\begin{align}
    H_{\rm ann}^{\mathfrak{su}(N),U} W_{\mathfrak{su}(N)}^U = 0, 
\end{align}
for $H_{\rm ann}^{\mathfrak{su}(N),U} = \sum_{j=1}^L (h_{\rm ann}^{\mathfrak{su}(N),U})_{j,j+1}$. 
Therefore, the frustration-free local construction of the annihilation term is preserved under the phase twist.

The same transformation can be applied to the tower-mixing perturbation introduced in Sec.~\ref{sec:tower_mixing}. The twisted mixing term is defined as
\begin{align}
    H_{\rm rhom}^{\mathfrak{su}(N),U}
    := U H_{\rm rhom}^{\mathfrak{su}(N)} U^\dagger.
\end{align}
Since $H_{\rm rhom}^{\mathfrak{su}(N)}$ is constructed from $\mathfrak{su}(N)$ root generators, $H_{\rm rhom}^{\mathfrak{su}(N),U}$ is correspondingly expressed in terms of the twisted root generators \eqref{eq:twist}. It therefore acts within the twisted invariant subspace, 
\begin{align}
    H_{\rm rhom}^{\mathfrak{su}(N),U} W_{\mathfrak{su}(N)}^U
    \subseteq W_{\mathfrak{su}(N)}^U,
\end{align}
and mixes the twisted tower states without destroying the algebraic closure.

In contrast to the root generators, the Cartan generators are invariant under the phase twist. Since $U_j$ is diagonal in the weight basis, 
\begin{align}
    U_j (h_a)_j U_j^\dagger = (h_a)_j,
\end{align}
and therefore
\begin{align}
    U H_{\rm tower}^{\mathfrak{su}(N)}U^\dagger
    = H_{\rm tower}^{\mathfrak{su}(N)}.
\end{align}
In the absence of the tower-mixing perturbation, the annihilation term vanishes within $W_{\mathfrak{su}(N)}^U$, while the Cartan term remains unchanged. Hence the twisted tower states 
\begin{align}
    |\Psi_{\bm n}^U\rangle
    := U|\Psi_{\bm n}\rangle
\end{align}
have exactly the same energies as their untwisted counterparts,
\begin{align}
    H_{\rm tower}^{\mathfrak{su}(N)} |\Psi_{\bm n}^U\rangle
    = E_{\bm n}|\Psi_{\bm n}^U\rangle,
\end{align}
where $E_{\bm n}$ is given by \eqref{eq:suN_energy}.
Thus, the lattice-like energy spectrum within the invariant subspace is unchanged by the phase twist.

In the presence of the tower-mixing perturbation, the Cartan term is no longer the only nonvanishing term within the invariant subspace. Nevertheless, the full twisted Hamiltonian is unitarily related to the untwisted one, 
\begin{align}
    H^{\mathfrak{su}(N),U} = U H^{\mathfrak{su}(N)} U^\dagger,
\end{align}
and therefore the energy spectrum within $W_{\mathfrak{su}(N)}^U$ is again identical to that within $W_{\mathfrak{su}(N)}$. The phase twist thus changes the local realization of the algebraically closed subspace and its generators without changing its spectral structure.

As a familiar example of the same phase-twisting mechanism, the
spin-$1$ $XY$ model discussed in Sec.~\ref{sec:XY} corresponds to the embedded
$\mathfrak{su}(2)$ construction with $k_1 = \pi$.

\section{Matrix-product state expression of the tower states} \label{app:MPS}

In this appendix, we explicitly provide the MPS expressions of the tower states \eqref{eq:PBW}.

\subsection{Original tower}

Using the matrix-product operator (MPO) representation~\cite{bib:MRB18} for the $n$th power of a translationally invariant sum of locally nilpotent single-site operators, the $n$th power of an $\mathfrak{su}(N)$ ladder operator $X$, whose local components are also nilpotent, is expressed as an MPO with bond dimension $n+1$:
\begin{align} \label{eq:MPO}
    &X^n = \Big( \sum_{j=1}^L X_j \Big)^n
    = n!\, \langle v_{\rm L}| \prod_{j=1}^L M_j |v_{\rm R} \rangle, \nonumber \\
    &(M_j)_{\mu,\nu} = \delta_{\mu,\nu}\, \bm{1}_j + \delta_{\mu+1,\nu}\, X_j.  
\end{align}
Note that the subscript $j$ labels the $j$th site of the spin chain. The boundary vectors, $(|v_{\rm L} \rangle)_{\rho} = \delta_{\rho,1}$ and $(|v_{\rm R} \rangle)_{\rho} = \delta_{\rho,n+1}$, fix the bond degrees of freedom. The indices $(\mu,\nu)$ label the matrix elements in the bond space.

The reference state $|\Psi_{\bm{0}} \rangle$ \eqref{eq:reference} of the tower states is simply given by the fully polarized direct-product state and hence has bond dimension one in the MPS representation. Taking into account that the tower states $|\Psi_{\bm{n}} \rangle$ are generated by successively applying the ordered ladder operators to the reference state, the state $|\Psi_{\bm{n}} \rangle$ admits an MPS representation with bond dimension $\prod_{a=1}^{N-1} (n_a+1)$. 

The same MPO-based construction \eqref{eq:MPO} can be applied to the linearly gauge-transformed $\mathfrak{su}(N)$ tower states generated by twisted ladder operators \eqref{eq:twist}, with the MPO tensor modified as
\begin{align}
    &(M_j^{\sigma,U})_{\mu,\nu} = e^{-i\sigma (k_a - k_{a-1}) (n+1-\mu)} \big(\delta_{\mu,\nu} \bm{1}_j + \delta_{\mu+1,\nu} (X^{\sigma}_a)_j \big), \qquad \sigma = \pm, 
\end{align}
without changing the bond dimension. Here we used the notation $X_a^+ = e_a$ and $X_a^- = f_a$. 

Using the known fact that the entanglement entropy of an MPS is bounded from above by the logarithm of its bond dimension, we conclude that the entanglement entropy of the tower states \eqref{eq:PBW} is bounded from above by 
\begin{align}
    S_{\rm EE}
    \leq \log \prod_{a=1}^{N-1} (n_a+1)
    = \sum_{a=1}^{N-1} \log(n_a+1).
\end{align}
This bound scales as $O(\log L)$ even in the finite-density regime, where $\sum_{a=1}^{N-1} n_a = O(L)$, and therefore obeys a sub-volume law. 

This derivation is the single-site nilpotent operator case of \cite{bib:OBCK20}.

\subsection{Tower-mixing case}

After adding a tower-mixing perturbation \eqref{eq:rhom_suN}, an $\mathfrak{su}(N)$ ladder operator and, consequently, the corresponding tower state are unitarily transformed by a uniform $\mathrm{SU}(N)$ rotation \eqref{eq:rotated_tower}. By this uniform $\mathrm{SU}(N)$ rotation $G^{\otimes L}$, an MPO expression of the ladder operator is transformed as 
\begin{align} 
    &(X^G)^n = \Big( \sum_{j=1}^L X^G_j \Big)^n
    = n!\, \langle v_{\rm L}| \prod_{j=1}^L M^G_j |v_{\rm R} \rangle, \nonumber \\
    &(M^G_j)_{\mu,\nu} = \delta_{\mu,\nu} \bm{1}_j + \delta_{\mu+1,\nu} X^G_j,   
\end{align}
where $X^G_j = G X_j G^{\dag}$, and hence its bond dimension does not change. 
The HWS $|\Psi_0 \rangle = |\lambda \rangle^{\otimes L}$ is transformed by the same $\mathrm{SU}(N)$ rotation as $|\Psi^G_0 \rangle = (G|\lambda \rangle)^{\otimes L}$, which remains a direct product state. 

Thus, the bond dimension of the MPS expression does not change by this uniform $\mathrm{SU}(N)$ rotation, implying that the entanglement entropy of a rotated tower state $|\Psi^G_{\bm{n}} \rangle = G^{\otimes L} |\Psi_{\bm{n}} \rangle$ is bounded from above by 
\begin{align}
    S^G_{\rm EE}
    \leq \log \prod_{a=1}^{N-1} (n_a+1)
    = \sum_{a=1}^{N-1} \log(n_a+1), 
\end{align}
which obeys a sub-volume law.

\end{appendix}

\bibliographystyle{SciPost_bibstyle}
\bibliography{references_doi_fixed}

\end{document}